# Learning-based Robust Speaker Counting and Separation with the Aid of Spatial Coherence

Yicheng Hsu[1] and Mingsian R. Bai[1,2]*

**Abstract**

A three-stage approach is proposed for speaker counting and speech separation in noisy and reverberant environments. In the spatial feature extraction, a spatial coherence matrix (SCM) is computed using whitened relative transfer functions (wRTFs) across time frames. The global activity functions of each speaker are estimated from a simplex constructed using the eigenvectors of the SCM, while the local coherence functions are computed from the coherence between the wRTFs of a time-frequency bin and the global activity function-weighted RTF of the target speaker. In speaker counting, we use the eigenvalues of the SCM and the maximum similarity of the interframe global activity distributions between two speakers as the input features to the speaker counting network (SCnet). In speaker separation, a global and local activity-driven network (GLADnet) is used to extract each independent speaker signal, which is particularly useful for highly overlapping speech signals. Experimental results obtained from the real meeting recordings show that the proposed system achieves superior speaker counting and speaker separation performance compared to previous publications without the prior knowledge of the array configurations.

## 1 Introduction

Blind speech separation (BSS) involves the extraction of individual speech sources from a mixed signal without prior knowledge of the speakers and mixing systems [1]. BSS finds application in smart voice assistants, hands-free teleconferencing, automatic meeting transcription, etc., where only mixed signals from single or multiple microphones are available. Several BSS algorithms have been developed based on different assumptions about the characteristics of the speech sources and the mixing systems [2]–[9]. Learning-based BSS approaches have recently received increased research attention due to advances in deep learning hardware and software. Promising results have been obtained using single-channel neural networks (NNs) [10]–[15]. To further improve separation performance, techniques have been developed that exploit the spatial information embedded in the microphone array signals began to emerge [16]–[19]. However, most of these BSS techniques assume a known number of speakers prior to separation. As a key step prior to speaker separation, speaker counting [20] is examined next.

Some studies have assumed the maximum number of speakers during speaker separation [21]–[24]. Another approach is to extract speech signals in a recursive manner [25]–[27], where the BSS problem has been tackled by a multi-pass source-extraction procedure based on a recurrent neural network (RNN). In contrast to the previous methods that use implicit speaker counting for separation, a multi-decoder DPRNN [28] uses a count-head to infer the number of speakers and multiple decoder heads to separate the signals. A speaker counting technique has been proposed using a scheme that alternates between speech enhancement and speaker separation [29]. Instead of exhaustive separation, one can selectively extract only the target speech signal, with the help of auxiliary information such as video images [30, 31], pre-enrolled utterances [32]–[34], and the location of the target speaker [35]–[38]. Although the target speaker extraction approach leads to significant performance improvements, the auxiliary information may not always be accessible. To overcome this problem, the speaker activity-driven speech extraction neural network [39] has been proposed to facilitate target speaker extraction by monitoring speaker activity. However, the speaker activity-driven speech extraction neural network is susceptible to adverse acoustic conditions in speaker extraction using speaker activity information alone. In such circumstances, multichannel approaches may be more advantageous than monochannel approaches. For example, deep clustering-based speaker counting and mask estimation have been incorporated into masking-based linear beamforming for speaker separation tasks [40]. Chazan et al. presented the use of a deep-neural-network (DNN)-based single-microphone concurrent speaker detector for source counting, followed by beamformer coefficient estimation for speaker separation [41, 42].

*Correspondence: msbai@pme.nthu.edu.tw

Despite the promising results obtained with DNN-based approaches, most network models require a large amount of data for training. Another limitation is that identical array configurations used in the test and training phases are preferred. Therefore, DSP-based approaches may have certain advantages [43]. Laufer-Goldshtein et al. proposed the global and local simplex separation algorithm by exploiting the correlation matrix of relative transfer functions (RTFs) across time frames [44]. The number of speakers is determined from the eigenvalue decay of the correlation matrix. The activity probabilities of each speaker are estimated from the simplex formed by the eigenvectors. In the separation stage, a spectral mask is computed for the identified dominant speakers, followed by spatial beamforming and post-filtering. Although the simplex-based approach is very effective in most cases, it does not work well for low-activity speakers [45].

In general, the DNN-based approaches show promise, but require extensive training data and could not generalize well to unseen array configurations. The DSP-based approaches require no training and often allow for low-resource implementation, but their performance depends on the array configuration. While the deep clustering-based speaker counting and mask estimation methods [40]–[42] are also array configuration-agnostic, their speaker counting capability relies on single-channel input feature, which can degrade counting performance in adverse acoustic conditions. Furthermore, the separation performance of these methods is dependent on the array configurations used.

The goal of this study is twofold. First, we reformulate a spatial feature that significantly improves the performance and robustness of source counting and separation. Second, we seek to leverage the strengths of DSP-based and learning-based methods for improved speaker counting and speaker separation performance, with robustness to unseen room impulse responses (RIRs) and array configurations. Inspired by the work of Gannot et al. [44, 46], which is a purely DSP-based approach, we propose a robust speaker counting and activity-driven speaker separation algorithm that combines statistical preprocessing and a neural network back-end. We formulate a modified spatial coherence matrix based on whitened relative transfer functions (wRTFs) as a spatial signature of directional sources. The whitening procedure provides spectrally rich phase information that proves to be a robust spatial signature for dealing with mismatched array configurations. In the speaker counting stage, our approach attempts to reliably estimate the number of active speakers in low-SNR and low-activity scenarios by incorporating eigenvalues from the spatial coherence matrix and the maximum similarity between the global activity distributions. In the speaker separation stage, the local coherence functions of each speaker are computed using the coherence between the wRTFs of each time-frequency (TF) bin and that weighted by the corresponding global activity function. The target masks for each speaker are estimated using a global and local activity-driven network (GLADnet), which remains effective for "mismatched" RIRs and array configurations not included in the training data.

We train our DNN models with RIRs simulated using the image-source method [47], while the trained models are tested using the measured RIRs recorded at Bar-Ilan University [48]. Real-life recordings from the LibriCSS meeting corpus [49] are also used to validate the proposed separation networks. In this study, the proposed speaker counting and speaker separation algorithms are compared with the simplex-based methods developed by Laufer-Goldshtein *et al.* [44] in terms of F1 scores and confusion matrices. Perceptual evaluation of speech quality (PESQ) [50] and word error rate (WER) are adopted as the performance measures in speaker separation tasks.

While inspired by Ref. [44], this study presents three main contributions that differ from the previous work. First, a learning-based robust speaker counting and activity-driven speaker separation algorithm is developed. Second, a modified spatial coherence matrix is formulated to effectively capture the spatial information of independent speakers. A novel idea based on the maximum similarity between the global activity distribution of two speakers over time frames is explored as an input feature for speaker counting. Third, an array configuration-agnostic GLADnet informed by the global and local speaker activities is proposed.

The remainder of this paper is organized as follows. Section 2 presents the problem formulation and a brief review of the simplex-based approach, which is used as the baseline in this study. Section 3 presents the proposed speaker counting and speaker separation system. In Section 4, we compare the proposed system with several baselines through extensive experiments. Section 5 concludes the paper.

## 2 Problem formulation and the baseline approach
### 2.1 Problem formulation

Consider a scenario in which the utterances of $J$ speakers are captured by $M$ distant microphones in a reverberant room. We assume that there is no prior knowledge of the array configuration. The array signal model is described in the short-time Fourier transform (STFT) domain. The received signal at the $m$th microphone can be written as

$$X^m(l,f) = \sum_{j=1}^{J} A_j^m(f) S_j(l,f) + V^m(l,f), \quad (1)$$

where $l$ and $f$ denote the time frame index and frequency bin index, respectively, $A_j^m(f)$ denotes the acoustic transfer function (ATF) between the $m$th microphone and the $j$th speaker, $S_j(l,f)$ denotes the signal of the $j$th speaker, and $V^m(l,f)$ denotes the additive sensor noise. This study aims to estimate the number of speakers $J$ (speaker counting) and extract independent speaker signals from the microphone mixture signals without information about the sources and the mixing process.

### 2.2 Baseline method: the simplex-based approach

In this section, we present the baseline by revisiting [44]. The simplex-based approach [44, 45] is based on the global and local simplex representations and relies on the assumption of the speech sparsity in the STFT domain [51]. By assuming speech sparsity, each TF bin is dominated by either the speaker or the noise. The ideal indicator selected in each TF bin can be expressed as

$$I_j(l,f) = \begin{cases} 1 & j\text{th speaker is dominant} \\ 0 & \text{otherwise} \end{cases} \quad (2)$$

If a TF bin is not dominated by any speakers, such a TF bin will be dominated by noise, i.e., $\sum_{j=1}^{J} I_j(l,f) = 0$. Let $p_j^G(l)$ be the global activity of speaker $j$ in frame $l$:

$$p_j^G(l) = \frac{1}{F} \sum_{f=1}^{F} I_j(l,f) \quad (3)$$

which is the global activity associated with the $j$th speaker in the $l$th frame. Note that the global activities $\{p_j^G(l)\}_{j=1}^{J}$ depend only on the frame index, not on the frequency index.

#### 2.2.1 Spatial feature extraction

Assuming speech sparsity in the TF domain, the relative transfer function (RTF) [52], which represents the ratio between the ATF of the $m$th microphone and the ATF of the first (reference) microphone, can be written as

$$R^m(l,k) = \frac{X^m(l,k)}{X^1(l,k)}$$

$$= \begin{cases} \dfrac{A_j^m(f)}{A_j^1(f)} & \text{for } I_j(l,f)=1,\ 1\le j\le J \\ \dfrac{V^m(l,f)}{V^1(l,f)} & \text{for } \sum_{j=1}^{J} I_j(l,f)=0 \end{cases} \quad (4)$$

In the following, a feature vector $\mathbf{r}(l)$ for each frame $l$ is defined to compose $D = 2\times(M-1)\times K$ elements of the real and imaginary parts of the computed ratios (4) for $1\le k\le K$ frequency bins and in ($M$-1) microphone signals:

$$\begin{aligned}\mathbf{r}^m(l) &= \begin{bmatrix} R^m(l,f_1) & R^m(l,f_2) & \cdots & R^m(l,f_K)\end{bmatrix} \\ \mathbf{r}^c(l) &= \begin{bmatrix} \mathbf{r}^2(l) & \mathbf{r}^3(l) & \cdots & \mathbf{r}^M(l)\end{bmatrix} \\ \mathbf{r}(l) &= \begin{bmatrix} \text{real}\{\mathbf{r}^c(l)\} & \text{imag}\{\mathbf{r}^c(l)\}\end{bmatrix}^T\end{aligned} \quad (5)$$

where $\{f_k\}_{k=1}^{K}$, are the selected frequencies. The correlation matrix $\mathbf{W} \in \mathbb{R}^{L\times L}$ is computed, where $[\mathbf{W}]_{ln} = \frac{1}{D}\mathbf{r}^T(l)\mathbf{r}(n)$. $\mathbf{W}$ can be approximated as [46]

$$\mathbf{W} \approx \mathbf{P}\mathbf{P}^T \quad (6)$$

where $\mathbf{P} = \begin{bmatrix} \mathbf{p}_1^G & \cdots & \mathbf{p}_J^G\end{bmatrix} \in \mathbb{R}^{L\times J}$ is composed of the global activity vectors $\mathbf{p}_j^G = \begin{bmatrix} p_j^G(1) & \cdots & p_j^G(L)\end{bmatrix}^T \in \mathbb{R}^{L\times 1}$ associated with the $j$th speaker.

### 2.2.2 Speaker counting

For $J$ independent speakers, the matrix $\mathbf{P}$ should have rank $J$. It follows that the number of speakers can be determined by counting the principal eigenvalues of the correlation matrix $\mathbf{W}$. However, selecting an appropriate threshold is not straightforward due to complex acoustic conditions. To select an appropriate threshold, the speaker counting problem has been formulated as a classification problem [44], where each class corresponds to a different number of speakers. A feature vector consisting of the first $J'$ principal eigenvalues of the correlation matrix is used as the input to the classifier

$$\mathbf{f}_{\text{baseline 1}} = \begin{bmatrix} \lambda_1 & \lambda_2 & \cdots & \lambda_{J'}\end{bmatrix}^T \quad (7)$$

where $J'$ is the maximum possible number of speakers and is set to 4 in this study. The multiclass support vector machine (SVM) is used as the classifier in [44].

### 2.2.3 Speaker separation

Once the number of speakers ($J$) is available, the eigenvectors associated with the $J$ largest eigenvalues for each frame $l$ are selected to form the global mapping vector

$$\mathbf{v}^G(l) = \begin{bmatrix} u_1(l), u_2(l), \ldots, u_J(l)\end{bmatrix}^T, \quad (8)$$

where $\{u_j(l)\}_{j=1}^{J}$, denotes the $l$th element associated with the $j$th eigenvector.

According to [44, 46], the global mapping vector $\mathbf{v}^G(l)$ can be expressed as a linear transformation of the global activity vector $\mathbf{p}^G(l)$:

$$\mathbf{v}^G(l) = \mathbf{G}\mathbf{p}^G(l) \quad (9)$$

with embedded information of speaker activities. The successive projection algorithm [53] can be applied to identify the simplex vertices and construct the transformation matrix $\mathbf{G} = [\mathbf{v}^G(l_1), \mathbf{v}^G(l_2), \ldots, \mathbf{v}^G(l_J)]$, where $\{l_j\}_{j=1}^{J}$ represent frame indices of the simplex vertices. Hence, the global activity can be computed.

$$\mathbf{p}^G(l) = [p_1^G(l), p_2^G(l), \ldots, p_J^G(l)]^T = \mathbf{G}^{-1}\mathbf{v}^G(l) \quad (10)$$

For the local mapping, each TF bin is assigned to a dominant speaker or noise. The spectral mask can be obtained by using the weighted nearest-neighbor rule.

$$M(l,f) = \underset{j\in\{1,\ldots,J+1\}}{\arg\max}\ \frac{1}{\pi_j}\sum_{n=1}^{L}\omega_{ln}(f)p_j^G(n) \quad (11)$$

where $\pi_j = \sum_{n=1}^{L} p_j^G(n)$ denotes the class normalization factor and $\omega_{ln}(f)$ is a Gaussian

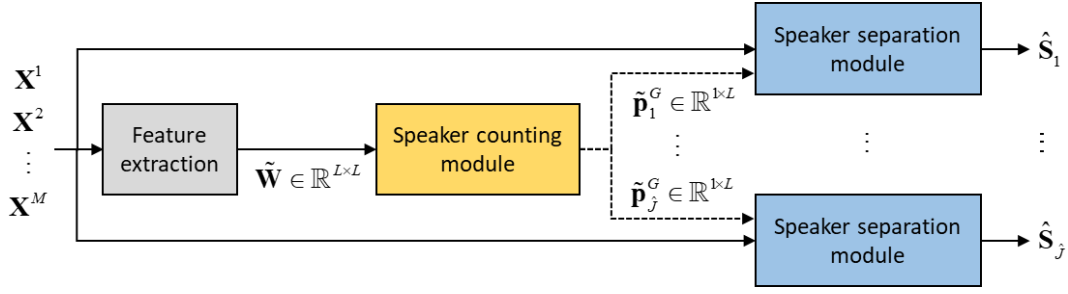

**Figure 1** Block diagram of the proposed speaker counting and separation system.

weighting function [34]:

$$\omega_{ln}(f) = \exp\{-\|\mathbf{r}(l,f) - \mathbf{r}(n,f)\|\} \quad (12)$$

that is inversely related to the distance in the space defined by the local representation $\{\mathbf{r}(l,f)\}_{l=1}^{L}$ between frame $n$ and frame $l$. The signal of the $j$th speaker can be estimated by applying the spectral mask in (11) to the reference microphone signal:

$$\hat{S}_j^{Mask}(l,f) = \begin{cases} X^1(l,f) & \text{if } M(l,f) = j \\ \beta X^1(l,f) & \text{otherwise,} \end{cases} \quad (13)$$

where $\beta$ is the attenuation factor to avoid musical noise. In this paper, $\beta$ is set to 0.1 as in [44].

A linearly constrained minimum variance (LCMV) beamformer can be used to extract each independent speaker signals [44, 45], with the weights below

$$\mathbf{w}_{LCMV} = \mathbf{R}_{nn}^{-1}(f)\hat{\mathbf{A}}(f)\left(\hat{\mathbf{A}}^H(f)\mathbf{R}_{nn}^{-1}(f)\hat{\mathbf{A}}(f)\right)^{-1}\mathbf{g}_j, \quad (14)$$

where $\hat{\mathbf{A}}(f) = [\hat{\mathbf{a}}_1(f), \ldots, \hat{\mathbf{a}}_J(f)]^T \in \mathbb{C}^{M \times J}$ denotes the RTF matrix with $\hat{\mathbf{a}}_j(f) = [\hat{A}_j^1(f), \hat{A}_j^2(f), \ldots, \hat{A}_j^M(f)]^T$ of the $j$th speaker and $\mathbf{R}_{nn}(f)$ is the noise covariance matrix. In this study, only sensor noise is assumed, i.e., $\mathbf{R}_{nn} = \sigma_{nn}\mathbf{I}$. As a result, (14) reduces to

$$\mathbf{w}_{LCMV} = \hat{\mathbf{A}}(f)\left(\hat{\mathbf{A}}^H(f)\hat{\mathbf{A}}(f)\right)^{-1}\mathbf{g}_j \quad (15)$$

where the RTF of the $j$th speaker can be estimated by

$$\hat{A}_j^m(f) = \frac{\sum_{l \in \mathcal{L}_j} X^m(l,f)X^{1*}(l,f)}{\sum_{l \in \mathcal{L}_j} X^1(l,f)X^{1*}(l,f)} \quad (16)$$

where $\mathcal{L}_j = \{l \mid p_j^G(l) > \varepsilon, l \in \{1, \ldots, L\}\}$ denotes the set of frames dominated by the $j$th speaker, and $\varepsilon = 0.2$ is an activity threshold.

To further illuminate the residual noise and interference, a single-channel mask is applied [44, 45], as given by

$$\hat{S}_j^{LCMV-Mask}(l,f) = \begin{cases} \mathbf{w}_{LCMV}^H \mathbf{x}(l,f)\mathbf{g}_j & \text{if } M(l,f) = j \\ \beta \mathbf{w}_{LCMV}^H \mathbf{x}(l,f)\mathbf{g}_j & \text{otherwise,} \end{cases} \quad (17)$$

where the vector $\mathbf{x}(l,f) = [X^1(l,f), \ldots, X^M(l,f)]^T$ denotes the microphone signals, $\mathbf{g}_j \in \mathbb{R}^{J \times 1}$ is a one-hot vector with one in the $j$th entry and zeros elsewhere, and $\beta = 0.2$ is a small factor to prevent from musical noise.

## 3 Proposed method

Inspired by the above simplex-based approach, we develop a robust speaker counting and separation system by exploiting spatial coherence features of array signals, as illustrated in Fig. 1. The system consists of three modules: the feature extraction module (Section 3.1), the speaker counting module (Section 3.2), and the speaker separation module (Section 3.3), as detailed in the sequel.

### 3.1 Spatial feature extraction

The simplex-based method [44] exploits the spatial information provided by the microphone array. As a result, spatial feature extraction plays a critical role in subsequent speaker counting and separation algorithms. Instead of the RTF used in [44], in this study we extract spatial information by whitening RTFs with no change in phase to enhance the spatial signature of the directional source, analogous to generalized cross-correlation with phase transformation (GCC-PHAT) [54]. In the light of the uncertainty principle [55], this helps to improve the time domain resolution for the computation of the spatial coherence matrix. Instead of the real feature vector used in the simplex-based approach, a "whitened" complex feature vector $\tilde{\mathbf{r}}(l)$ is defined as

$$\tilde{\mathbf{r}}(l) = [\tilde{\mathbf{r}}(l, f_1) \ \tilde{\mathbf{r}}(l, f_2) \ \cdots \ \tilde{\mathbf{r}}(l, f_K)]^T \in \mathbb{C}^{(M-1)K \times 1} \quad (18)$$

where

$$\tilde{\mathbf{r}}(l, f) = \left[ \frac{R^2(l,f)}{|R^2(l,f)|} \ \cdots \ \frac{R^M(l,f)}{|R^M(l,f)|} \right]$$

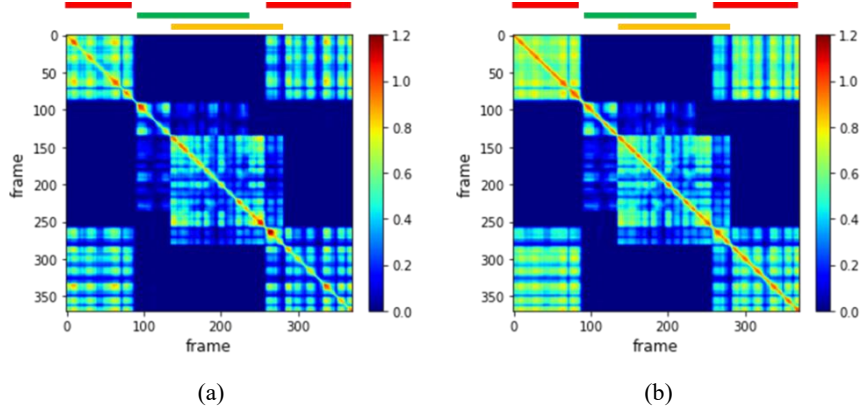

**Figure 2** Examples of (a) the spatial correlation matrix $\mathbf{W}$ and (b) the spatial coherence matrix $\tilde{\mathbf{W}}$. The color bars at the top of each figure indicate the active span of each speaker.

$R^m(l, f)$ is defined in (4), and $\{f_k\}_{k=1}^{K}$ is the selected frequency band as in (5). Next, we construct a spatial coherence matrix $\tilde{\mathbf{W}} \in \mathbb{R}^{L \times L}$ with the $ln$th entry defined as

$$\tilde{W}_{ln} = \frac{\operatorname{Re}\{\tilde{\mathbf{r}}^H(l)\tilde{\mathbf{r}}(n)\}}{\|\tilde{\mathbf{r}}(l)\|\|\tilde{\mathbf{r}}(n)\|} = \frac{1}{\tilde{D}}\operatorname{Re}\{\tilde{\mathbf{r}}^H(l)\tilde{\mathbf{r}}(n)\} \quad (19)$$

where $\operatorname{Re}\{\cdot\}$ is the real-part operator, $\|\cdot\|$ denotes the $l_2$-norm, and $\tilde{D} = \|\tilde{\mathbf{r}}(l)\|\|\tilde{\mathbf{r}}(n)\| = (M-1)K$ due to the fact that the feature vectors have been whitened. Note that the complex inner product of $\tilde{\mathbf{r}}(l)$ and $\tilde{\mathbf{r}}(n)$ is computed, which can also be regarded as a sign-sensitive cosine similarity based on the Euclidean angle [56]. An example of the spatial correlation matrix computed using the method reported in the references [44]–[46] and the proposed spatial coherence matrix are compared in Fig. 2, which is generated using a 12-second clip with a three-speaker mixture captured by an eight-element uniform linear array (ULA) with interelement spacing of 8 cm. The image in Fig. 2(b) is preferable to Fig. 2(a) because the time span of the proposed spatial coherence matrix aligns better than the baseline, especially at the overlap, as shown by the ground-truth activity bar at the top of the figure. This suggests that the proposed spatial coherence matrix is effective in capturing speaker activity, much like a voice activity detector. In addition, the range of the proposed coherence matrix is within [-1, 1], which is a desired property for network training.

### 3.2 Speaker counting

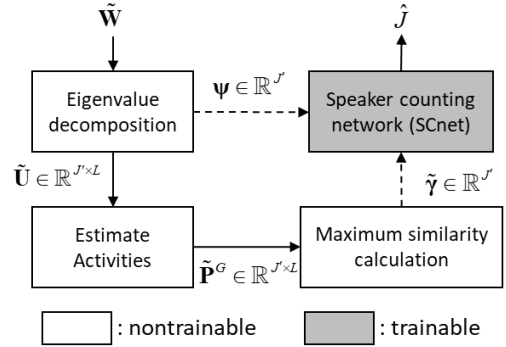

**Figure 3** Flowchart of the proposed speaker counting approach.

The flowchart of the proposed speaker counting approach is detailed in Fig. 3. Two features related to the speaker count are extracted from the spatial coherence matrix $\tilde{\mathbf{W}}$ and input to the speaker counting network

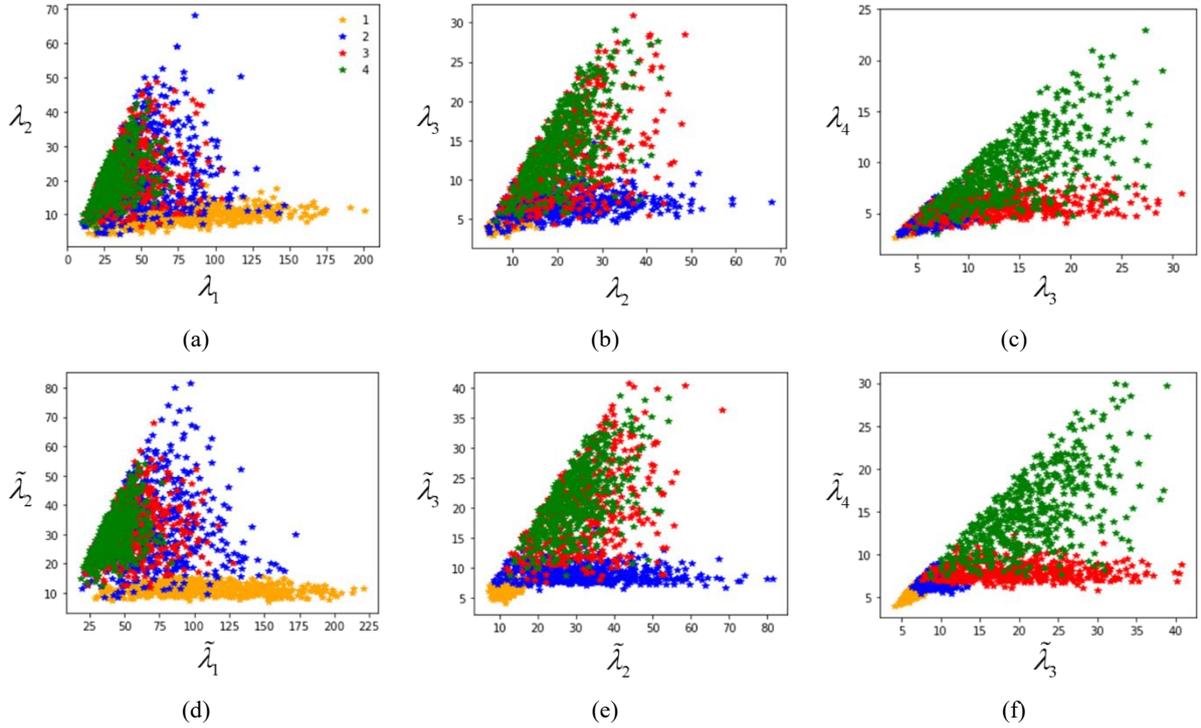

**Figure 4** Scatter plots of the eigenvalues corresponding to the observations with $J \in \{1, 2, 3, 4\}$ speakers. Each cross with different color represents an observation corresponding to different number of speakers. The upper row shows the result with the correlation matrix $\mathbf{W}$ and the lower row is the result with the coherent matrix $\tilde{\mathbf{W}}$.

(SCnet), as will be detailed next.

In this study, we propose to use the eigenvalues $\{\tilde{\lambda}_n\}_{n=1}^{L}$ of the spatial coherence matrix $\tilde{\mathbf{W}}$ as the feature for the classifier. An example of scatter pattern of the eigenvalues to discriminate between different speaker count classes, $J \in \{1, 2, 3, 4\}$, is illustrated in Fig. 4. We generated 2000-sample speech mixtures for 1-4 speakers, with 0%, 10%, 20%, 30%, 40% overlap ratios. Sensor noise was added with 10 dB SNR. Dry signals were convoluted with the measured RIRs selected from the Multi-Channel Impulse Responses Database [48] that was recorded using an eight-element ULA with interelement spacing of 8 cm and T60 = 0.61 s. Each cross in the figure represents one observation to specify the number of speakers. Figure 4 shows the ability of the eigenvalues obtained from the correlation matrix and the coherence matrix to discriminate between different numbers of speakers. In addition, the eigenvalues of the coherence matrix $\tilde{\mathbf{W}}$ can discriminate between different numbers of speakers better than those of the correlation matrix $\mathbf{W}$. However, some of the observations cannot be classified into the correct class according to the eigenvalues. In this study, we evaluate the similarity between global activities as auxiliary information to address the cases where the principal eigenvalue-based counting method does not work.

Apart from eigenvalues of the spatial coherence matrix, another feature that can help speaker counting is introduced to deal with meeting scenarios in which the overlap ratio of conversation is often less than 20% [57]. For such scenarios, we first calculate a similarity matrix $\tilde{\boldsymbol{\gamma}}^j \in \mathbb{R}^{j \times j}$ of the first $j$ global activities with the $pq$-th entry defined as

$$\tilde{\gamma}_{pq}^j = \frac{\tilde{\mathbf{p}}_p^G \cdot \tilde{\mathbf{p}}_q^G}{\|\tilde{\mathbf{p}}_p^G\| \|\tilde{\mathbf{p}}_q^G\|} \quad (20)$$

where "·" denotes the inner product, $\tilde{\mathbf{p}}_p^G \in \mathbb{R}^{L \times 1}$ and $\tilde{\mathbf{p}}_q^G \in \mathbb{R}^{L \times 1}$ denote the $p$th and $q$th global activities estimated from the spatial coherence matrix $\tilde{\mathbf{W}}$, and $1 \leq p, q \leq j$. Next, we find the maximum similarity value of all entries but the diagonal entries.

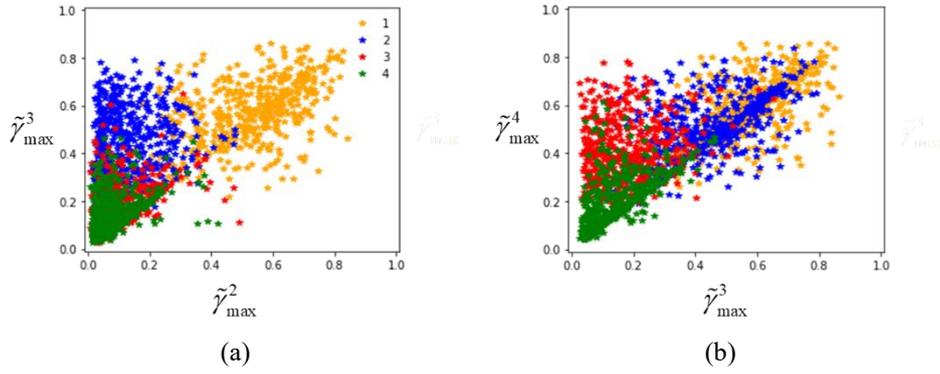

**Figure 5** Scatter plots of the maximum similarity to the observations with $J \in \{1, 2, 3, 4\}$ speakers. Each cross with different color represents an observation corresponding to different number of speakers.

$$\tilde{\gamma}_{max}^j = \max_{p,q} \left( \tilde{\boldsymbol{\gamma}}^j - \mathbf{I} \right)_{pq} \tag{21}$$

Similarly, $\gamma_{max}^j$ denotes the maximum similarity calculated using the first $j$ global activities obtained from the spatial correlation matrix $\mathbf{W}$. An example of scatter pattern of the maximum similarity to discriminate between different speaker count classes, $J \in \{1, 2, 3, 4\}$, is illustrated in Fig. 5. The data generation is identical to those of Fig. 4. To visualize the separability by using the proposed feature, we plot the scatter plot by the projection onto a two-dimensional feature space. Figure 5 suggests that the observations are separable by the maximum similarity, which helps to classify the number of speakers. In Fig. 5(a), the single-speaker observations and the two to four speaker observations are clearly separable along the $\tilde{\gamma}_{max}^2$ coordinate. The one or two speaker observations and the three or four speaker observations are clearly separable along the $\tilde{\gamma}_{max}^3$ coordinate. In Fig. 5(b), the one to three speaker observations and the four speaker observations are clearly separable along the $\tilde{\gamma}_{max}^4$ coordinate.

In this study, the speaker counting problem is formulated as a classification problem as in Ref. [44] with four classes corresponding to 1 to 4 speakers. For each observation (audio clip), the number of speakers is indicated by a one-hot vector $\mathbf{z} \in \mathbb{R}^{4 \times 1}$. For inference, the prediction is the highest probability of the output distribution. Three different input feature vectors are defined for the assessment of speaker counting performance:

$$\begin{aligned} \mathbf{f}_{baseline\,2} &= \left[ \frac{\lambda_2}{\lambda_1} \cdots \frac{\lambda_{J'}}{\lambda_1} \; \gamma_{max}^2 \cdots \gamma_{max}^{J'} \right]^T \in \mathbb{R}^{2(J'-1)} \\ \mathbf{f}_{proposal\,1} &= \left[ \frac{\tilde{\lambda}_2}{\tilde{\lambda}_1} \cdots \frac{\tilde{\lambda}_{J'}}{\tilde{\lambda}_1} \right]^T \in \mathbb{R}^{J'-1} \\ \mathbf{f}_{proposal\,2} &= \left[ \frac{\tilde{\lambda}_2}{\tilde{\lambda}_1} \cdots \frac{\tilde{\lambda}_{J'}}{\tilde{\lambda}_1} \; \tilde{\gamma}_{max}^2 \cdots \tilde{\gamma}_{max}^{J'} \right]^T \in \mathbb{R}^{2(J'-1)}, \end{aligned} \tag{22}$$

where $J' = 4$ is the maximum possible number of speakers, and the eigenvalues are normalized by the maximum eigenvalue to improve convergence. Features $\mathbf{f}_{baseline\,2}$ is obtained from the spatial correlation matrix $\mathbf{W}$, whereas features $\mathbf{f}_{proposal\,1}$ and $\mathbf{f}_{proposal\,2}$ are obtained from the proposed spatial coherence matrix $\tilde{\mathbf{W}}$.

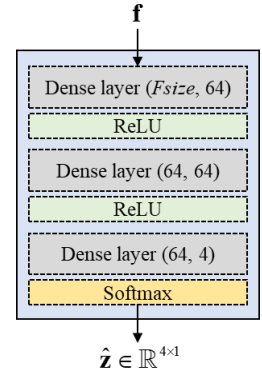

**Figure 6** Speaker counting network (SCnet).

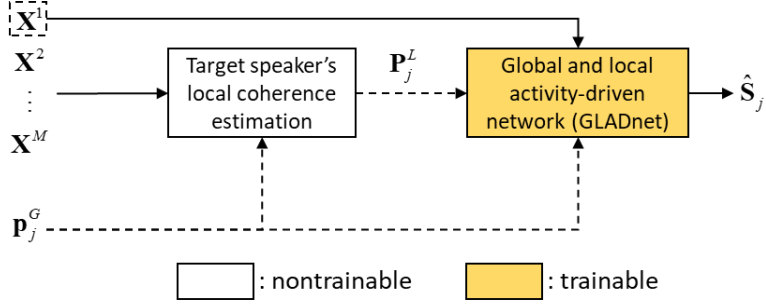

**Figure 7** Block diagram of the proposed speaker separation module.

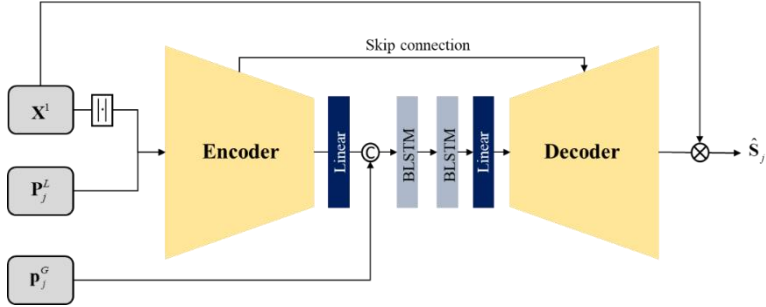

**Figure 8** The GLADnet.

A DNN model termed SCnet is used as the classifier for speaker counting. Figure 6 shows an SCnet consisting of three dense layers followed by a rectified linear unit (ReLU) activation, with softmax activation in the output layer. In addition, ($F_{\text{size}}$,64) means a dense layer with input size = $F_{\text{size}}$ and output size = 64. The cross-entropy is used as the loss function in network training.

### 3.3 Speaker separation

The simplex-based method relies solely on the spatial cue to perform the subsequent beamforming, which depends on the specific array configuration. In contrast, our learning-based approach uses global and local spatial activity features to train the model, as shown in Fig. 7. The proposed system consists of two main modules: 1) the local coherence estimation of independent speakers, which monitors the local activity of each speaker according to the global activity of the speaker, and 2) the global and local activity-driven network (GLADnet), which extracts the speaker signal with the auxiliary information about the global and local activities of the speaker.

In the local coherence estimation of a speaker, the local coherence is calculated between the wRTF of the target speaker and the wRTF of each TF bin. The wRTF of the $j$th speaker is calculated as

$$\tilde{\mathbf{a}}_j(f) = \left[ \frac{\hat{A}_j^2(f)}{|\hat{A}_j^2(f)|} \quad \cdots \quad \frac{\hat{A}_j^M(f)}{|\hat{A}_j^M(f)|} \right]^T \quad (23)$$

where $\hat{A}_j^m(f)$ is the estimated RTF. Thus, the local coherence of the $j$th speaker can be calculated

$$\begin{aligned} p_j^L(l,f) &= \frac{\text{Re}\{\tilde{\mathbf{a}}_j^H(l)\tilde{\mathbf{r}}(l,f)\}}{\|\tilde{\mathbf{a}}_j(l)\|\|\tilde{\mathbf{r}}(l,f)\|} \\ &= \frac{1}{M-1}\text{Re}\{\tilde{\mathbf{a}}_j^H(l)\tilde{\mathbf{r}}(l,f)\}, \end{aligned} \quad (24)$$

where $\tilde{\mathbf{r}}(l,f)$ is given by the equation under (14). Local coherence serves to inform the DNN about the local activity of a speaker.

GLADnet is based on a convolutional recurrent network [58], as illustrated in Fig. 8. The network has three inputs: the magnitude spectrogram of the reference microphone signal, the global activity of the speaker, and the local activity of the speaker. GLADnet has six symmetric encoder and decoder layers with an 8-16-32-128-128-128 filter. The convolutional blocks feature a separable convolution layer, followed by batch normalization, and exponential linear unit activation. The output layer terminates with sigmoid activation. The convolution kernel and step size are set to (3,2) and (2,1), respectively. Note that 1 × 1 pathway convolutions (PConv) are used as skip connections, which leads to considerable parameter reduction with little performance degradation. The global activity is concatenated to the output of the linear layer with 256 nodes in each time frame. The resulting vector is then fed to the following bidirectional long short-term memory layers with 256 nodes to sift out the latent features pertaining to each speaker. The soft mask estimated by the network is multiplied element-wise with the noisy magnitude spectrogram to yield an enhanced spectrogram. The complete complex spectrogram can be obtained by combining the enhanced magnitude spectrogram with the phase of the noisy spectrogram. The network is trained to minimize the compressed mean square error between the masked magnitude ($\hat{\mathbf{S}}$) and the ground-truth magnitude ($\mathbf{S}$)

$$J_{CMSE} = \sum_{t,f} \left\| |\mathbf{S}|^c - |\hat{\mathbf{S}}|^c \right\|_F^2 \qquad (25)$$

where $c = 0.3$ is the compression factor and $\|\ \|_F$ denotes the Frobenius norm.

## 4 Experimental study

Experiments were performed to validate the proposed learning-based speaker counting and separation system. The networks were trained on the simulated RIRs and tested on the measured RIRs with different T60s and array configurations recorded at Bar-Ilan University [48]. For meeting scenarios, we also tested the proposed system on real meeting recordings from the LibriCSS meeting corpus [49].

### 4.1 Training and validation dataset

In total, 50,000 and 5000 samples were used in training and validation, respectively. Dry speech signals selected from the *train-clean-360* subset of the LibriSpeech corpus [59] were used for training and validation. Noisy speech mixtures edited in 12-s clips were prepared with different numbers of speakers $J \in \{1, 2, 3, 4\}$ in reverberation conditions and signal-to-noise ratios (SNRs) between −5 dB and 5 dB. The overlap ratio of the speech mixtures varied from 0% to 40%. Reverberant microphone signals were simulated by filtering the dry signals with the simulated RIRs using the image-source method [47]. The reverberation time was within the range of [0.2, 0.6] s. Sensor noise was added with SNR = 15, 25, and 35 dB. In this study, simulated (Gaussian) noise was used to simulate the sensor noise. Two microphone array geometries were used for training and validation, as depicted in Fig. 9. The first microphone array is an eight-element ULA with interelement spacing of 8 cm. The geometry of the second array is similar to that of the seven-element uniform circular array (UCA) used in the LibriCSS dataset [49] which has one microphone at the center and the other six uniformly distributed around a circle with a radius of 4.25 cm. The RIRs of rectangular rooms with randomly generated dimensions (length, width, and height) in the range of [3 × 3 × 2.5, 7 × 7 × 3] m were simulated. The ULA was placed at 0.5 m from the wall, while the UCA was placed at the center of the room. Any two speakers were separated by at least 15°.

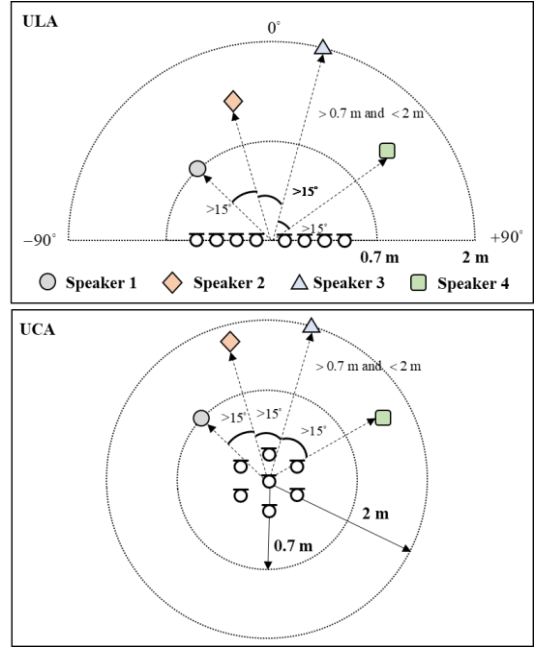

**Figure 9** Settings for network training with different microphone array geometries.

## 4.2 Implementation and evaluation metrics

In this study, the signal frame was 128 ms long with a 32 ms stride. A 2048-point fast Fourier transform was used. The sample rate was 16 kHz. The feature vectors in (5) and (18) comprised $K = 257$ frequency bins in 1-3 kHz. We chose this frequency range because, as in Ref. [44], it performed well in all of the scenarios examined for different simulated and measured RIRs and array configurations. In the experiment, SCnet and GLADnet were trained using the Adam optimizer with a learning rate of 0.001 and a gradient norm clipping of 3. The learning rate was halved if the validation loss did not improve for three consecutive epochs.

The F1 score and the confusion matrix are used to evaluate the speaker counting performance. The F1 score is a measure of the accuracy of a test in classification problems. It is defined as the harmonic mean of precision and recall [60]. PESQ [50] is used as a metric for speech quality and is computed only in the period when the speech is present. In addition, we also evaluate the WER achieved by the proposed system compared to the baselines, by using a transformer-based pre-trained model from the SpeechBrain toolkit [61]. The pre-trained model was trained on the LibriSpeech dataset. The WER obtained with this model when tested on the *test-clean* subset is 1.9%.

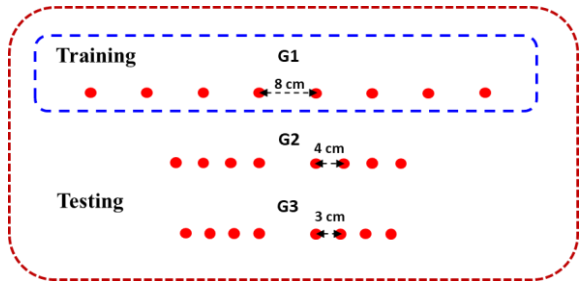

**Figure 10** Microphone array settings for experiments to investigate the effects of array configurations.

## 4.3 Spatial feature robustness

In this section, we aim to investigate the robustness of the algorithm with respect to the spatial correlation matrix and the spatial coherence matrix for measured RIRs and unseen array geometries. The proposed spatial coherence matrix based on wRTFs is used as a spatial signature for directional sources. The whitening process provides spectrally rich information that better accommodates unseen array configurations and measured RIRs. To see this, we compute the Modal Assurance Criterion (MAC) value on the spatial correlation matrix and the spatial coherence matrix for various unseen array configurations and RIRs. First, we vectorize the spatial matrix as $\psi = [\mathbf{w}_1 \ \mathbf{w}_2 \ \ldots \ \mathbf{w}_L]^T \in \mathbb{R}^{L^2 \times 1}$, where $\mathbf{w}_l = [W_{l1} \ W_{l2} \ \cdots \ W_{lL}] \in \mathbb{R}^{L \times 1}$. $\psi$ and $\psi'$ represent feature vectors associated with two spatial matrices. The MAC value between $\psi$ and $\psi'$ is defined as

$$MAC(\psi, \psi') = \frac{(\psi^T \psi')^2}{\psi^T \psi \psi'^T \psi'} \qquad (26)$$

To evaluate the robustness of the proposed spatial feature extraction method, we generated four different test datasets, each consisting of 500 samples. The first three datasets (G1, G2, and G3) were generated using measured RIRs from the Multi-Channel Impulse Responses Database [48], while the last dataset (sG1) was generated using simulated RIRs. As shown in Fig. 10, the first array configuration (G1) is included in the training set, while the second and third array configurations (G2 and G3) are considered "unseen" to the trained model. Note that sG1 had the same array configuration as G1, but with simulated RIRs. Tables 1 and 2 summary the MAC values obtained using the spatial correlation matrix and spatial coherence matrix. The off-diagonal MAC values of the spatial coherence matrix are consistently close to one and larger than those of the spatial correlation matrix. The MAC test demonstrates that the proposed spatial coherence matrix exhibits superior robustness to different array configurations and RIRs compared to the spatial correlation matrix. This property is desirable for the subsequent learning-based speaker counting and speaker separation approaches when dealing with unseen array configurations and measured RIRs.

Table 1 MAC values calculated using the spatial correlation matrix for various array configurations and RIRs

| MAC | sG1 | G1 | G2 | G3 |
|---|---|---|---|---|
| sG1 | 1 | 0.923 | 0.918 | 0.906 |
| G1 | 0.923 | 1 | 0.918 | 0.910 |
| G2 | 0.918 | 0.918 | 1 | 0.919 |
| G3 | 0.906 | 0.910 | 0.919 | 1 |

Table 2 MAC values calculated using the spatial coherence matrix for various array configurations and RIRs

| MAC | sG1 | G1 | G2 | G3 |
|---|---|---|---|---|
| sG1 | 1 | 0.989 | 0.986 | 0.978 |
| G1 | 0.989 | 1 | 0.986 | 0.975 |
| G2 | 0.986 | 0.986 | 1 | 0.993 |
| G3 | 0.978 | 0.975 | 0.993 | 1 |

### 4.4 Speaker counting performance

In the following, we examine several speaker counting methods for different levels of sensor noise and T60s. We generated 2000-sample speech mixtures for 1-4 speakers, with 0%, 10%, 20%, 30%, 40% overlap ratios, and dry speech signals from the *test-clean* subset of the LibriSpeech corpus. Sensor noise was added with SNR = 10, 20, and 30 dB. The measured RIRs were selected from the Multi-channel Impulse Responses Database [48] recorded using an eight-element ULA with interelement spacing of 8 cm and T60 = 0.36, 0.61 s at Bar-Ilan University. The RIRs were measured in 15° intervals from −90° to 90° at distances of 1 and 2 m from the array center. Table 3 summarizes the speaker counting results in F1 scores. We compare the proposed counting approaches with two baselines. Baseline 1 is the method proposed in [44]. The SVM classifier with $\mathbf{f}_{\text{baseline 1}}$ in (7) as the input feature is used for training. Baseline 2 is the SCnet trained with $\mathbf{f}_{\text{baseline 2}}$ in (22). For the proposed methods, proposals 1 and 2 represent the SCnet trained with $\mathbf{f}_{\text{proposal 1}}$ and $\mathbf{f}_{\text{proposal 2}}$ in (22). The speaker counting performance summarized in Table 3 suggests that baseline 1 performs comparably with baseline 2 in high SNR conditions. However, the speaker counting performance of baseline 1 degrades significantly as the SNR decreases. The feature using the eigenvalues obtained from the spatial coherence matrix (proposal 1) significantly outperform those obtained from the spatial correlation matrix (baseline 1), especially when the SNR is low. In addition, the method trained with the maximum similarity (proposal 2) could further improve the speaker counting performance over the method trained with eigenvalues only (proposal 1). In this study, speaker counting is highly dependent on the quality of spatial information extracted from the microphone array. However, it should be noted that spatial features tend to degrade as the SNR decreases. As a result, the counting performance may be relatively lower at SNR = 10.

Next, we investigate speaker counting in low-activity scenarios using four-speaker mixtures, where the first speaker was active for only 5% of the time. In Table 4, we see a significant performance degradation in the SCnet trained on the eigenvalues of the spatial correlation matrix (baseline 1), even in high-SNR conditions. In contrast, the SCnet trained on the eigenvalues and the maximum similarities computed using the proposed spatial coherence matrix (proposal 2) performs quite satisfactorily despite the unbalanced speaker activity.

Lastly, we investigate speaker counting using the real-life recordings from the LibriCSS dataset [49]. There are 10 one-hour sessions, including six 10-min mini-sessions in each session with different speaker overlap ratios (0S, 0L, 10%, 20%, 30%, and 40%). In the 0% case, 0S and 0L represent the signals with short and long silence periods, where inter-utterance silence lasts between 0.1-0.5s and 2.9-3.0s. The test data was pre-segmented into 12-second clips containing 1 to 4 speakers in each session. The speaker count of each audio clip was labeled by using the ground-truth information. In addition, the dataset contains 511, 1119, 614, and 154 examples for one, two, three, and four speakers, respectively. The results of speaker counting are summarized in the confusion matrices depicted in Fig. 11. The F1 scores for the baselines 1 and 2, proposals 1 and 2 were 88.37%, 92.44%, 96.48%, and 97.36%. From Fig. 11, we can see that the methods trained on the features from the spatial coherence matrix (proposals 1 and 2) outperform the methods trained on the features from the spatial correlation matrix (baselines 1 and 2). Figures 11(c) and (d) show that the methods trained on maximum similarities (proposal 2) yield significantly lower underestimation rates than the methods trained on eigenvalues only (proposal 1). For the BSS problems, underestimation can undermine the subsequent separation, while overestimation is less critical. In summary, we extract spatial information by whitening the RTFs without changing the phase to enhance the spatial signature of the directional source, analogous to generalized cross-correlation with phase transformation (GCC-PHAT) [54]. In the light of the uncertainty principle [55], this helps to improve the time domain resolution for the computation of the spatial coherence matrix, which in turn leads to a more accurate estimation of the spatial activity, especially in low SNR cases. This enables a more accurate estimation of the maximum similarity of two global activities as independent activities, without overlooking scenarios with low activity speakers.

**Table 3** Comparison of speaker sounting performance under different acoustical conditions in terms of F1 score

| T60 (ms) | 360 | | | 610 | | | |
|---|---|---|---|---|---|---|---|
| SNR (dB) | 30 | 20 | 10 | 30 | 20 | 10 | Avg. |
| baseline 1 | 99.40 | 94.42 | 55.94 | 98.81 | 93.72 | 59.00 | 83.55 |
| baseline 2 | 99.52 | 96.22 | 82.53 | 99.57 | 96.54 | 84.94 | 93.22 |
| proposal 1 | 99.62 | 98.66 | 90.79 | 99.72 | 98.63 | 91.29 | 96.45 |
| proposal 2 | **99.75** | **99.37** | **91.01** | **99.75** | **99.25** | **91.88** | **96.84** |

**Table 4** Comparsion of low-activity speaker counting performance under different acoustical conditions in terms of F1 score

| T60 (ms) | 360 | | | 610 | | | |
|---|---|---|---|---|---|---|---|
| SNR (dB) | 30 | 20 | 10 | 30 | 20 | 10 | Avg. |
| baseline 1 | 91.34 | 85.91 | 54.31 | 92.50 | 84.92 | 50.73 | 77.22 |
| baseline 2 | 97.65 | 89.49 | 64.27 | 96.29 | 86.47 | 65.78 | 82.73 |
| proposal 1 | 99.70 | 95.58 | 71.17 | 99.16 | 94.41 | 74.47 | 89.08 |
| proposal 2 | **99.70** | **98.43** | **78.21** | **99.75** | **98.10** | **80.22** | **92.40** |

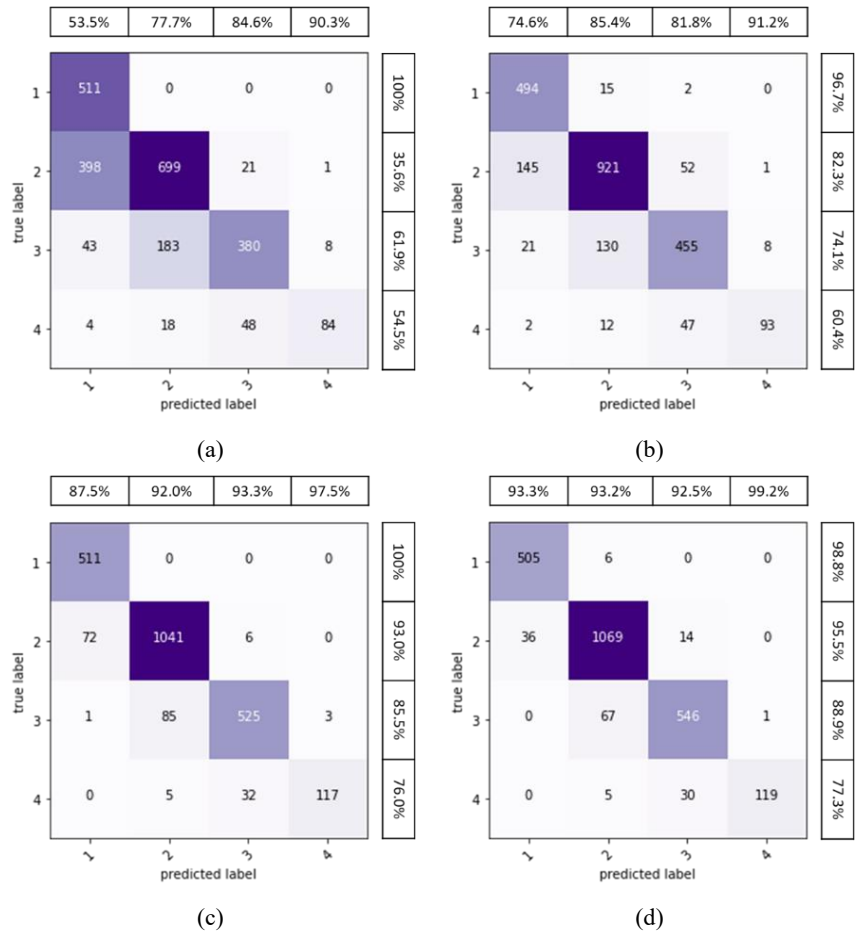

**Figure 11** Confusion matrices for the speaker counting results obtained using (a) baseline 1, (b) baseline 2, (c) proposal 1, and (d) proposal 2.

Furthermore, unlike most multichannel source counting methods, which typically require more microphones than sources, the simplex-based and the proposed methods are limited by the total number of frames used to compute the spatial correlation matrix and the spatial coherence matrix, not the number of microphones. This implies that, in theory, there is virtually no limit to the number of speakers that can be identified. In fact, the only limit on counting accuracy is the degree of time overlap. To see this, we give two examples with different speaker activity patterns to show the maximum number of independent speakers that can be identified using ULAs with 2-5 elements evenly spaced at 8 cm.

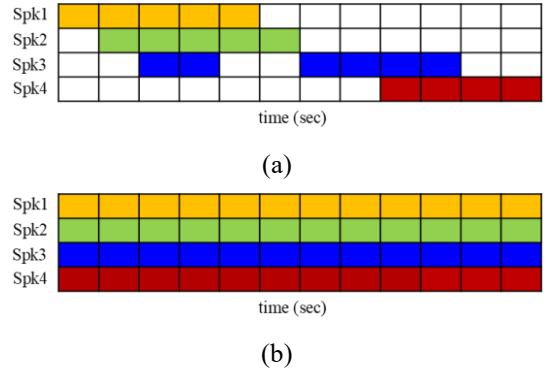

**Figure 12** Ground truth speaker activities for (a) Case I and (b) Case II.

Case I represents a scenario where four speakers are active in moderately overlapping time periods, as shown in Fig. 12(a). Note that at 2-4 seconds, three speakers are active concurrently. Inspection of Fig. 13(a) indicates that the spatial coherence matrices associated with different numbers of microphones remain very similar. In this case, the eigenvalue distribution analysis reveals that the number of sources can be accurately estimated, even when the number of speakers (4) exceeds the number of microphones (5), as shown in Fig. 14(a).

Case II presents a scenario where the proposed source counting method fails, where four independent speakers are active with 100% overlap, as shown in Fig. 12(b). In this case, the spatial coherence matrices in Fig. 13(b) show no meaningful patterns of activity, regardless of the number of microphones. The eigenvalue distribution analysis in Fig. 14(b) provides an incorrect estimate, one. In summary, methods based on simplex preprocessing are not limited by the number of microphones, but rather by the overlap percentage of the speaker activity time span.

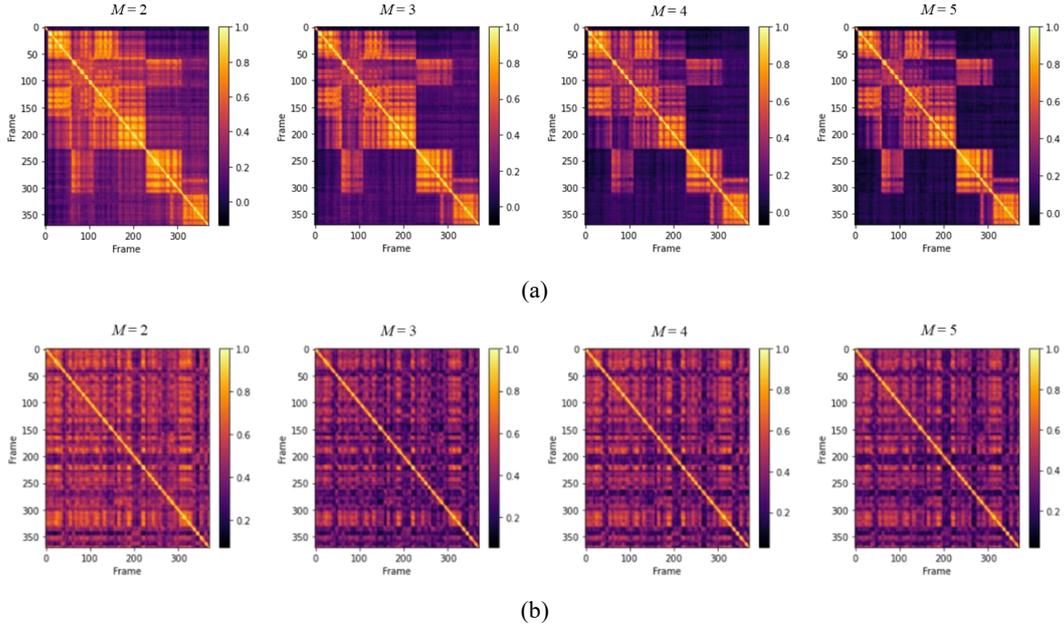

(a)

(b)

**Figure 13** Spatial coherence matrices for different number of microphones in (a) Case I and (b) Case II.

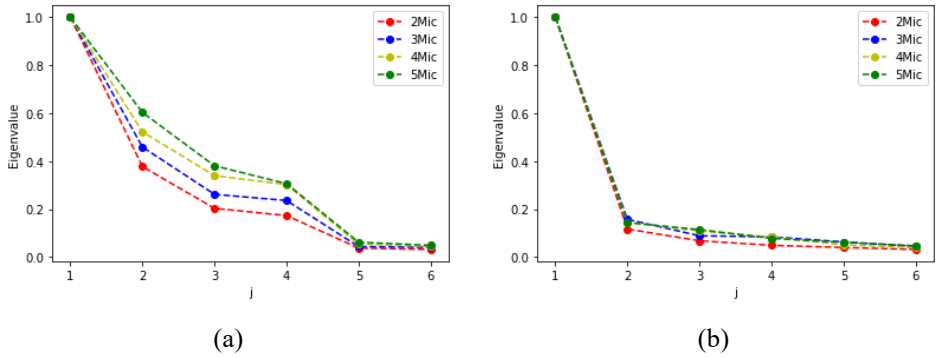

(a)     (b)

**Figure 14** Eigenvalue distribution in descending order of the spatial coherence matrix for (a) Case I and (b) Case II

### 4.5 Speaker separation performance

In the following, we compare the proposed speaker separation approach (GLADnet) with three baselines. The first baseline (Mask) uses only a spectral mask (13). The second baseline (LCMV-Mask) is the simplex-based approach [44, 45] with beamforming and spectral masking (17). The third is the GLADnet, which is trained only on the global activity, called the global activity-driven network (GADnet). To evaluate the robustness of the proposed speaker separation approach when applied to unseen RIRs and array configurations, we created three 2000-sample test datasets for three different array configurations (G1, G2, and G3) using the measured RIRs from the Multi-Channel Impulse Responses Database [48]. The array configurations G1, G2, and G3 are shown in Fig. 10.

First, we examine the separation performance using the G1 configuration for different overlap ratios and T60s. The results in Fig. 15 show that the proposed GLADnet outperforms the three baselines in terms of speech quality. The performance of the GADnet, which is not trained with spatial features, degrades drastically as the overlap ratio increases. While the LCMV-Mask method achieves comparable WER to GLADnet at moderate T60 = 360 ms, its separation performance drops sharply at high reverberation.

Next, the effect of array configurations on separation performance is investigated. Figure 16 reveals that the speech quality (PESQ) and the ASR performance (WER) using the LCMV-Mask method degrade as the array spacing and the array aperture decrease, even for moderate T60's. In contrast, the proposed GLADnet performs quite satisfactorily despite the unseen RIRs and array geometries.

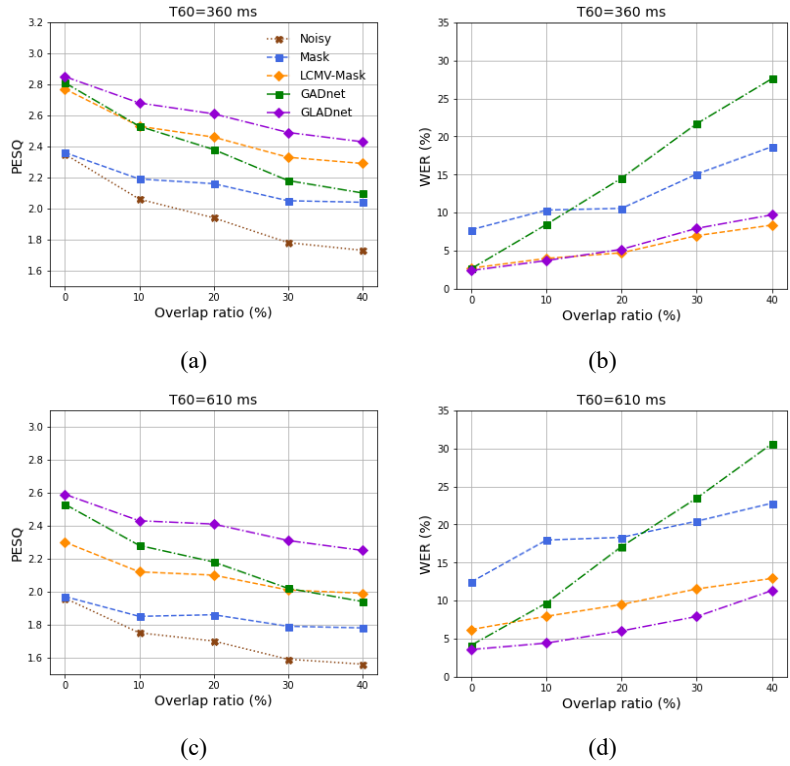

**Figure 15** Comparison of separation performance with array configuration (G1) in terms of (a), (c) PESQ and (b), (d) WER for different overlap ratios.

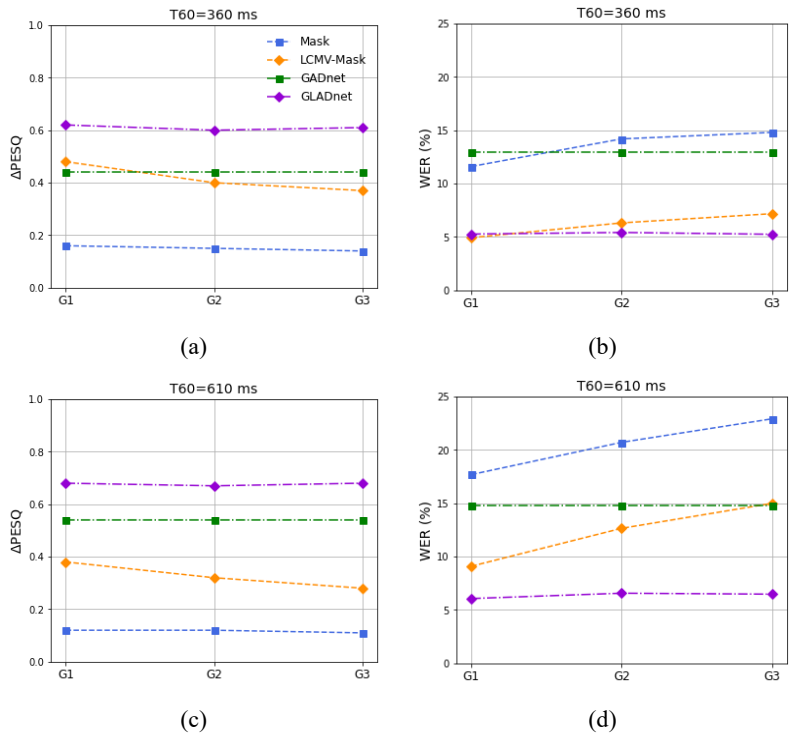

**Figure 16** Comparison of separation performance with array configurations (G1, G2, and G3) in terms of (a), (c) PESQ and (b), (d) WER for different array configurations.

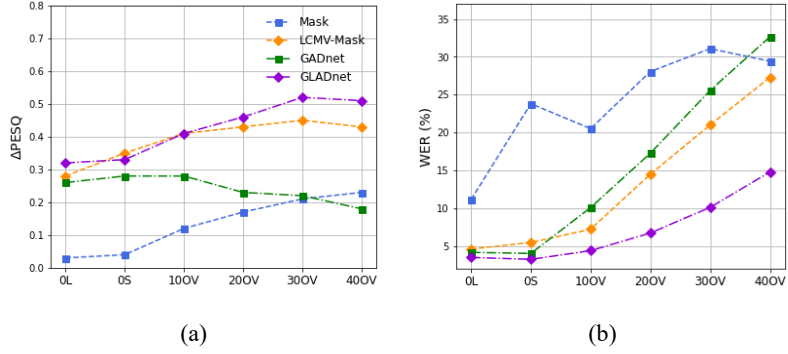

(a)                                      (b)

**Figure 17** Comparison of separation performance in terms of (a) PESQ and (b) WER for the LibriCSS dataset.

We also evaluated the proposed network in speaker separation using a more realistic LibriCSS dataset. The dataset generation for network testing is identical to that for speaker counting. Figure 17 shows that the LCMV-Mask method has a comparable performance to the proposed GLADnet when the overlap ratio is low. However, the performance of the LCMV-Mask drops dramatically at high overlap ratios. In addition, GADnet performs satisfactorily only for non-overlapping speech mixtures. In summary, the separation performance of baselines such as Mask and LCMV-Mask, which rely solely on spatial information, can be significantly affected by the inter-element spacing and array aperture. On the other hand, the baseline GADnet, which relies solely on spectral information, can suffer performance degradation in adverse acoustic conditions such as large reverberation and high overlap ratios. In contrast to these baselines, the proposed GLADnet exploits both spatial and spectral information to achieve superior performance in terms of PESQ and WER metrics. In addition, the GLADnet is trained using the global and local activities derived from the wRTFs, which is less sensitive to unseen RIRs and array configurations.

## 5 Conclusions

In this paper, a learning-based robust speaker counting and separation system has been implemented by integrating array signal processing and DNN. In feature extraction, the spatial coherence matrix computed with wRTFs across time frames shows superior robustness to different array configurations and RIRs compared to the spatial correlation matrix. In speaker counting, the SCnet trained on the eigenvalues and the maximum similarities obtained from the spatial coherence matrix is conducive to speaker counting in adverse acoustic conditions, especially in unbalanced voice activity scenarios. In speaker separation, the GLADnet based on global and local spatial activities proves to be capable of effective and robust enhancement with different overlap ratios for unseen RIRs and array configurations, which is highly desirable for real-world applications.